\documentclass[11pt,twoside]{article}
\usepackage{asp2010}

\resetcounters

\begin{document}

\title{Improving resolution and depth of astronomical observations via modern mathematical methods for image analysis}
\author{Marco Castellano$^1$, Daniele Ottaviani$^2$, Adriano Fontana$^1$,  Emiliano Merlin$^1$, Stefano Pilo$^1$, Maurizio Falcone$^2$
\affil{$^1$INAF - Osservatorio Astronomico di Roma, Via Frascati 33, I--00040, Monteporzio, Italy.}
\affil{$^2$Dipartimento di Matematica, Sapienza Universit\`{a} di Roma, P.le Aldo Moro, 2 - 00185 Roma, Italy}
}

\begin{abstract}
In the past years modern mathematical methods for image analysis have led to a revolution in many fields, from computer vision to scientific imaging. However, some recently developed image processing techniques successfully exploited by other sectors have been rarely, if ever, experimented on astronomical observations. We present here tests of two classes of variational image enhancement techniques: "structure-texture decomposition" and "super-resolution" showing that they are effective in improving the quality of observations. Structure-texture decomposition allows to recover faint sources previously hidden by the background noise, effectively increasing the depth of available observations. Super-resolution yields an higher-resolution and a better sampled image out of a set of low resolution frames, thus mitigating problematics in data analysis arising from the difference in resolution/sampling between different instruments, as in the case of EUCLID VIS and NIR imagers.
\end{abstract}

\section{Structure-texture image decomposition}
A general approach to the denoising problem is based on the assumption that an image $f$ can be regarded as composed of a structural part $u$, (i.e. the  objects in the image), and a textural part $v$ which corresponds to finest details plus the noise. In this paper we have considered image decomposition models based on total variation regularization methods following the approach described in \citet{Aujol}.
\begin{figure}[!ht]
\center
\includegraphics[scale=0.2]{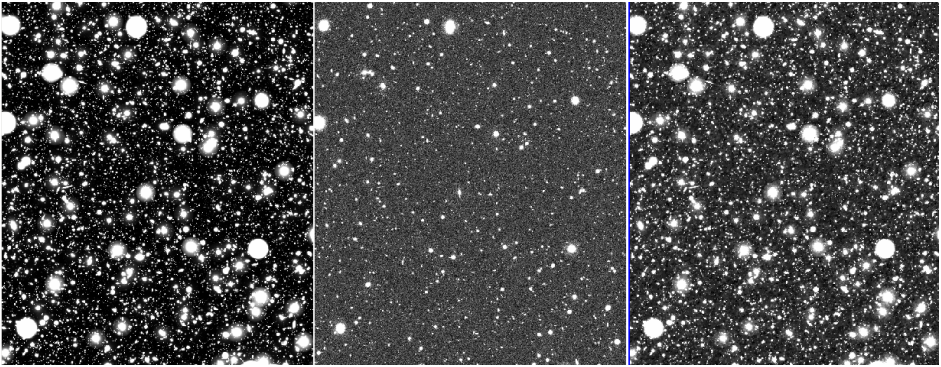}
\caption{From left to right: simulated noise-free mosaic, simulated observed mosaic (CANDELS-GOODS depth), "Structure" image obtained thorugh TVL2 decomposition.}\label{fig1}
\end{figure}
Such image decomposition technique is based on the minimization of a functional with two terms, one based on the total variation and a second one on a different norm adapted to the texture component.
Given an image $f$ defined in a set $\Omega$, and being $BV(\Omega)$ the space of functions with limited total variation in $\Omega$ (i.e. $V(f,\Omega)=\int\limits_\Omega\left|\nabla f(x)\right|\mathrm{d}x< +\infty$) we can decompose $f$ into its two componentes by minimizing:
\begin{equation}\label{denoisingP}
\inf \left(\int \left|Du\right| + \lambda  \left \| v \right \|^p_X\right),
\end{equation}
where $\left \| . \right \|^p_X$ indicates the norm of a given space $X$ (e.g. $L^2(\Omega)$) and the minimum is found among all functions $(u,v)\in\left(BV(\Omega)\times X\right)$ such that $u+v=f$. The parameter $p$ is a natural exponent, and $\lambda$ is the so-called \textit{splitting parameter}. The best decomposition is found at the $\lambda$ for which the correlation between $u$ and $v$ reaches a minimum. 
We have designed a C++ code, named \verb|ATVD|  (Astro-Total Variation Denoiser) which implements three versions of the technique, based respectively on the TV-L$^2$ ($X=L^2(\Omega)$), TV-L$^1$ ($X=L^1(\Omega)$) and TVG ($X$ being a Banach space as defined in \citet{Aujol}) norms.

\begin{figure}[!ht]
\center
\includegraphics[scale=0.28]{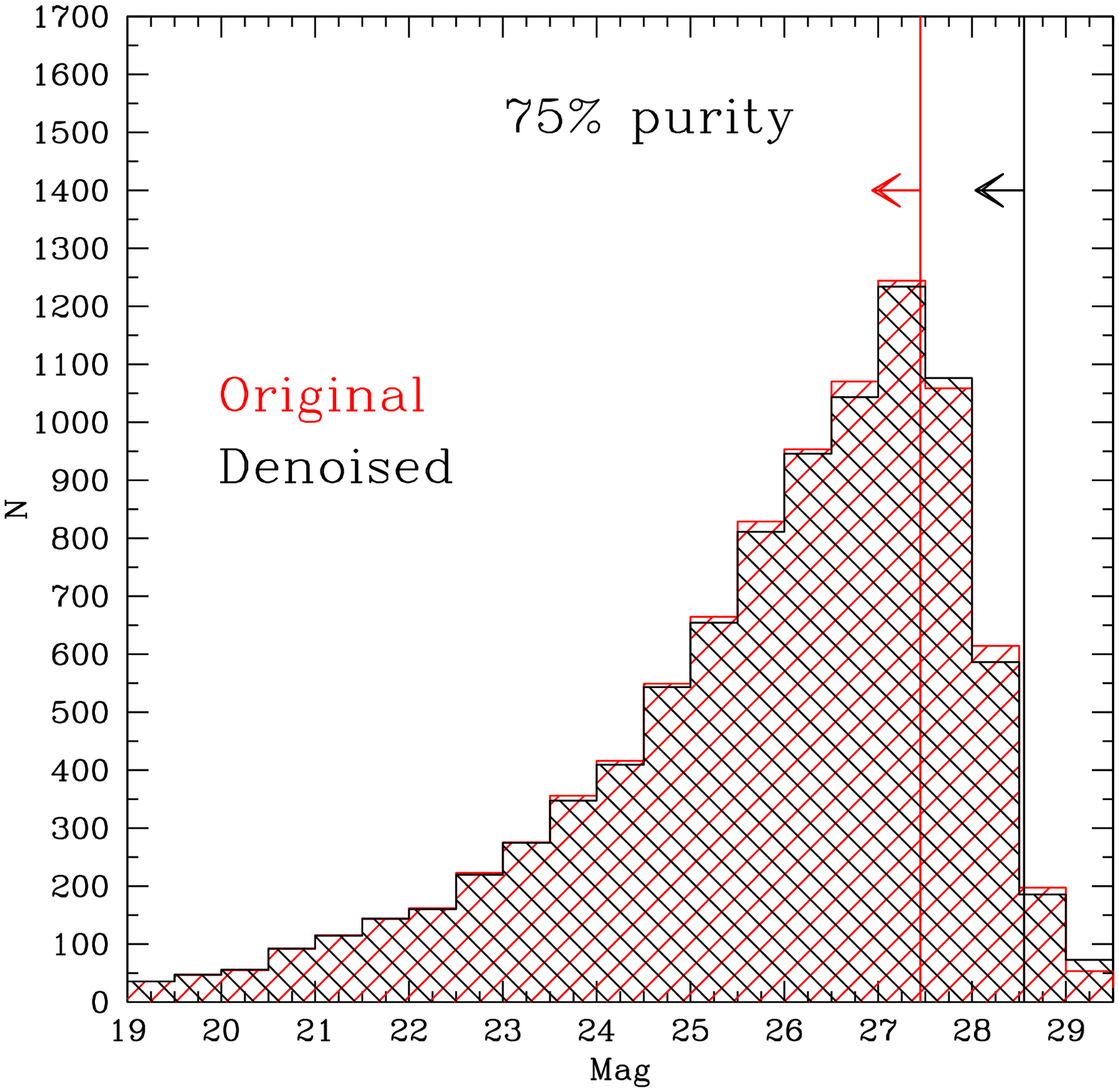}
\includegraphics[scale=0.28]{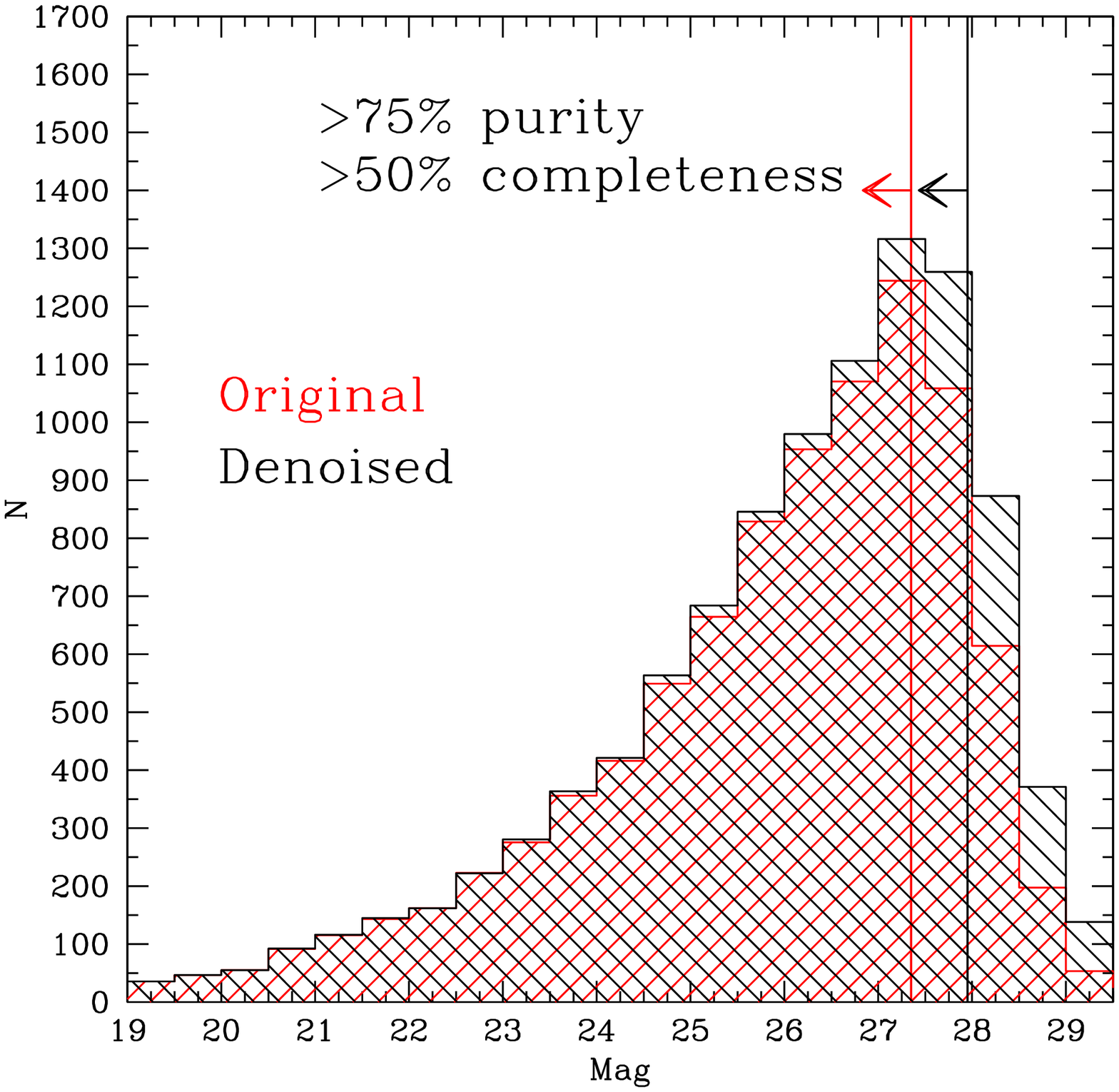}
\includegraphics[scale=0.26]{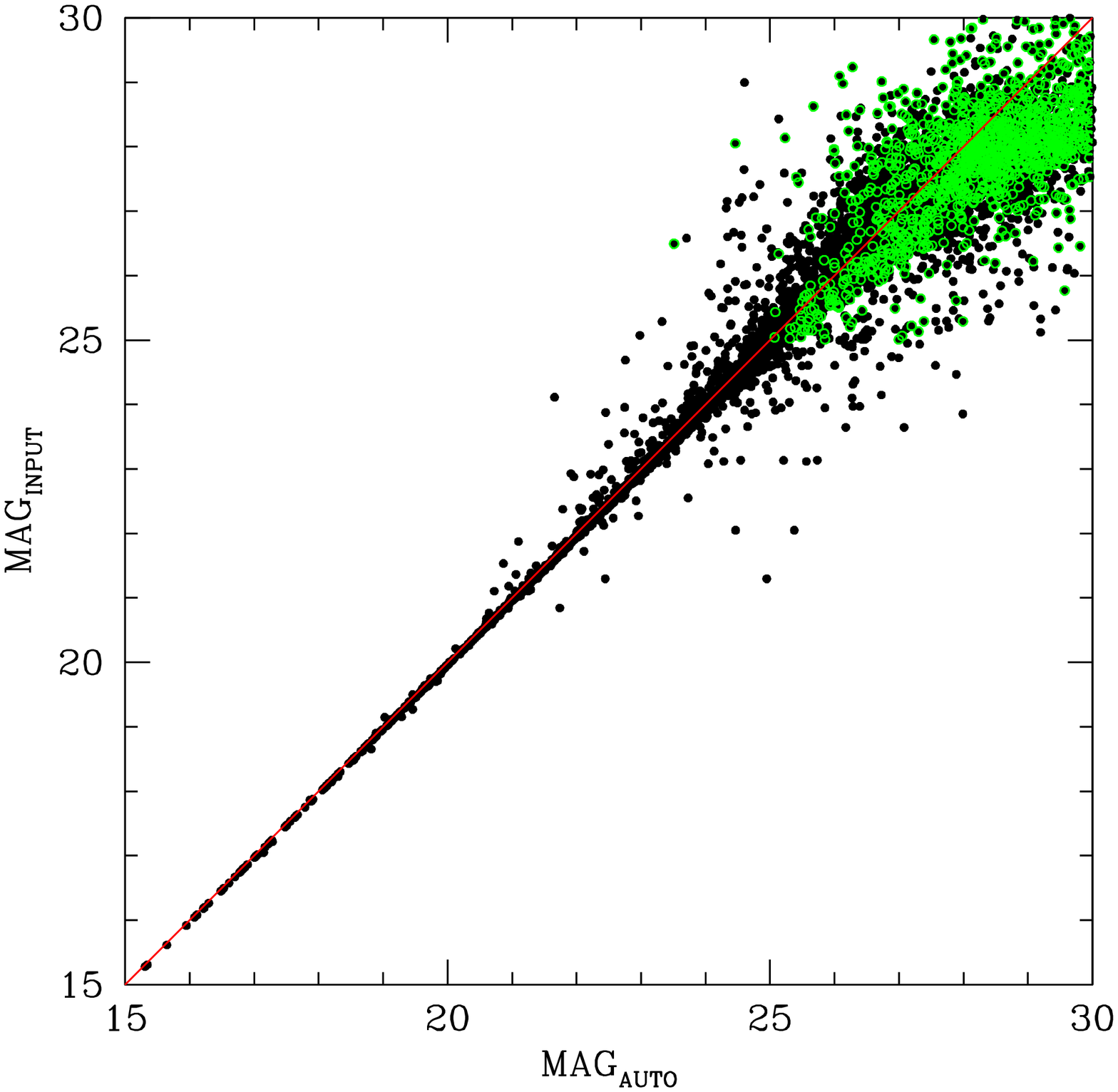}
\includegraphics[scale=0.22]{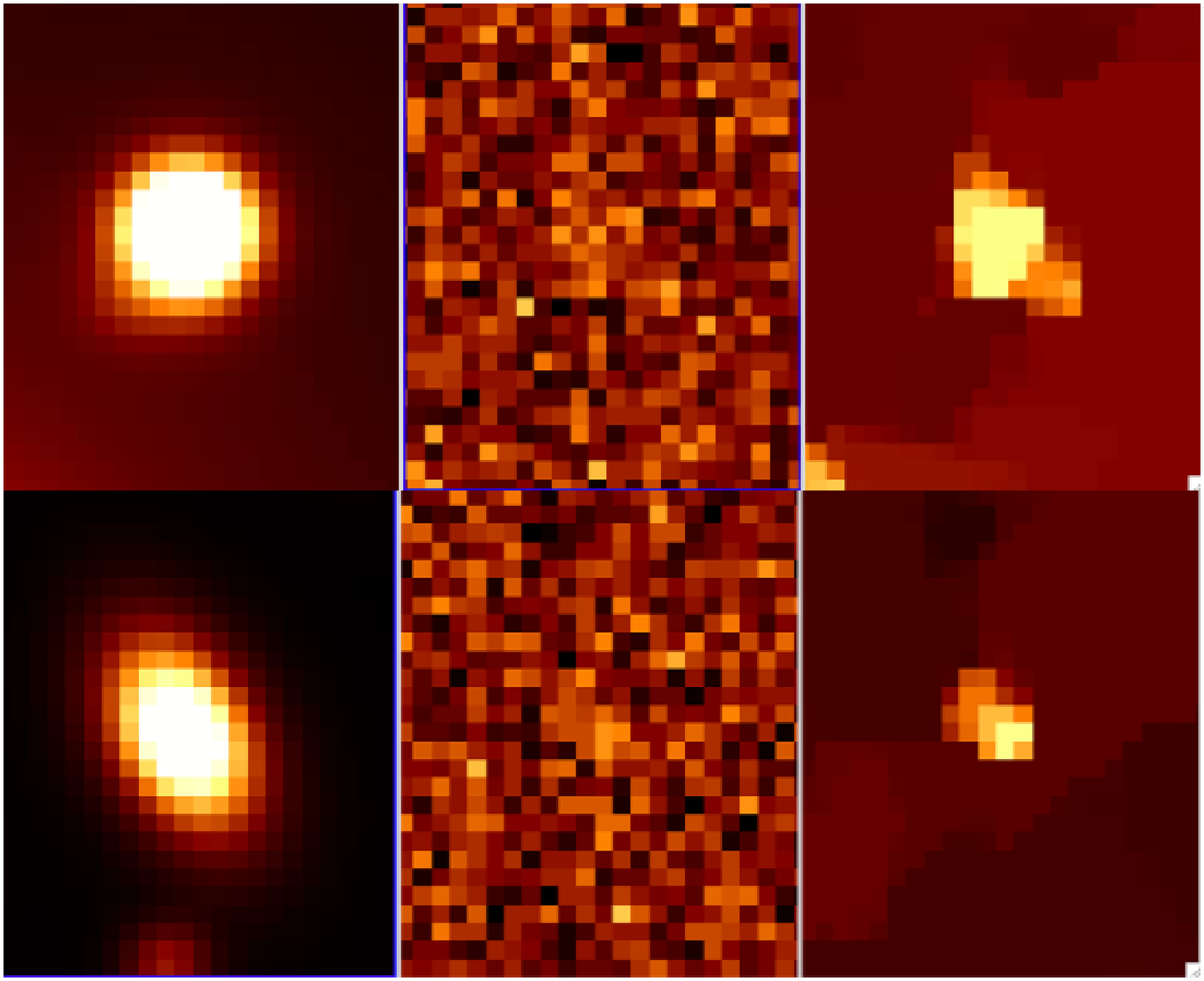}
\caption{Top panels: magnitude counts on the original and on the denoised (Structure) image maximising purity (left) or completeness (right) on the latter. Bottom panels: on the left, comparison between magnitudes measured on the denoised image and input magnitudes (additional faint sources not found on the noisy mosaic are shown in green). Right: sources detected on the Structure image but not in the input noisy one as they appear in the noise-free simulated mosaic, in the noisy image, and in the Structure image.}\label{fig2}
\end{figure}

\textbf{Tests on simulated HST images}. We have tested the performance of the decomposition technique method on a 15$\times$15 arcmin simulated image reproducing the average depth of the F160W (H160 hereafter) CANDELS GOODS-South mosaic \citep{Grogin2011}. The analysis of a simulated image in which the position of sources below and above then nominal detection limit are known, allows to test the effectiveness of decomposition techniques at increasing the depth of astronomical observations. We decomposed the simulated image through our ATVD code using different norms and we found that the TV-L$^2$ one is the most effective. The best Structure image (H160$_{St}$ hereafter) obtained is shown in Fig.\ref{fig1} together with the input noisy one (H160$_{inp}$) and, as a reference, the simulated image without noise.
At a visual inspection the resulting Structure image appears deeper and richer in details than the input observed mosaic.  To quantitatively assess the advantages of the TV-L$^2$ decomposition we have run SExtractor \citep{bertin} on both  H160$_{St}$ and H160$_{inp}$. By varying SExtractor input parameters affecting source detection we have estimated \textit{completeness} and \textit{purity} of the extracted catalogues. A comparison of the best detection performances on  H160$_{St}$ and H160$_{inp}$ is shown in Fig.\ref{fig2}. Source extraction on the Structure component H160$_{St}$ yields to an higher purity of the catalogue ($\sim$0.5 mags deeper confidence threshold) at a similar completeness reached on the original H160$_{inp}$. Alternatively, by pushing detection to lower S/N, we can obtain a much higher completeness with acceptable contamination levels (right panel in Fig.\ref{fig2}) with respect to H160$_{inp}$.
As shown in Fig.\ref{fig2} magnitudes estimated on H160$_{St}$ are reliable also for those sources up to mag$\sim$30 which are undetectable in the original mosaic.
\section{Super-resolution}

\begin{figure}[!ht]
\center
\includegraphics[scale=0.2]{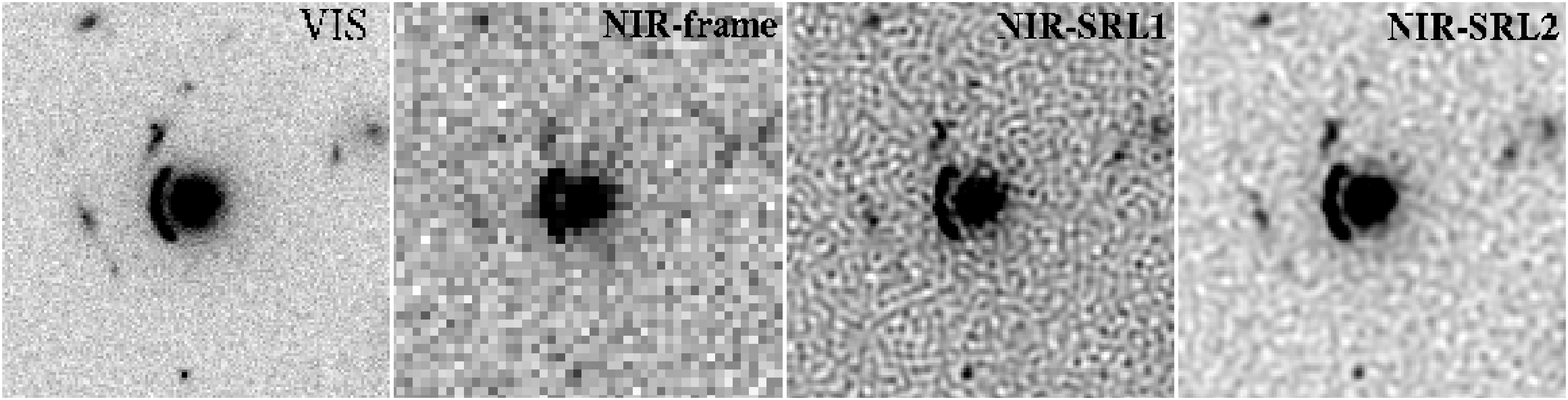}
\includegraphics[scale=0.34]{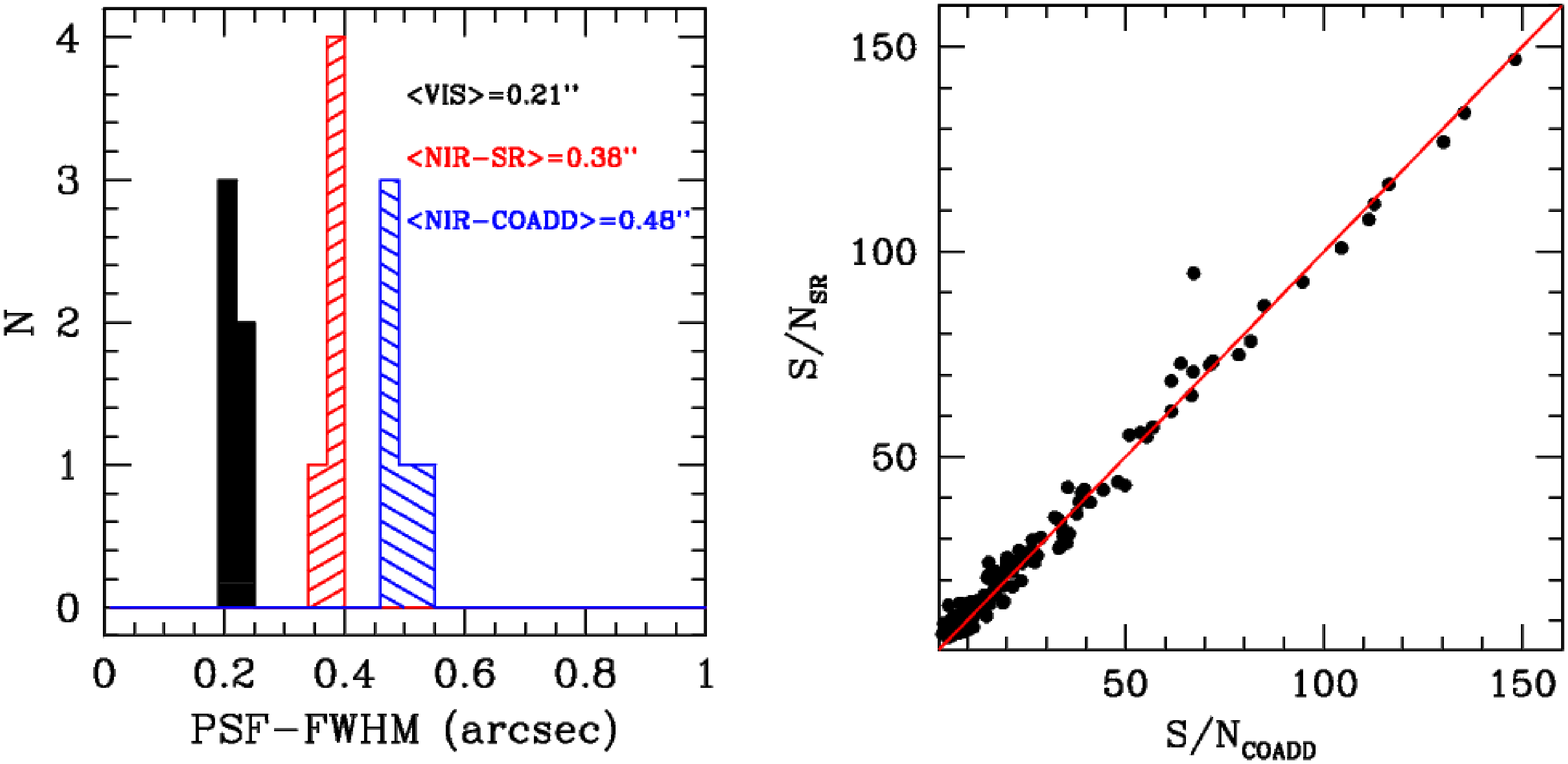}
\caption{Top: simulation of a VIS image, one NIR frame, super-resolved NIR TV-L$^1$ and TV-L$^2$ images obtained from 9 NIR frames. Bottom left: PSF-FWHM measured on: the VIS image, the SR-NIR at the VIS pixel-scale, and on a NIR coadded mosaic resampled to the VIS pixel-scale. Right: comparison between the S/N of sources in the SR-TVL2 mosaic and in a standard coadded NIR mosaic.}\label{fig3}
\end{figure}
A low-resolution, blurred frame can be considered as the result of the application of a number of operators (downsampling $D$, warping $W$, convolution $H$), to an ideal high resolution image, plus the addition of a noise component ($\underline{\epsilon}$) \citep[e.g.][]{irani1991improving}:
$\underline{x}_{L}=D H W\underline{x}_{H}+\underline{\epsilon}=K\underline{x}_{H}+\underline{\epsilon}$.
Following this image formation model, it can be shown that given N low resolution frames with pixel sampling $z$, we can recover an higher resolution well-sampled image with pixel-sampling $z/\sqrt(N)$ by minimizing the following energy function:
\begin{equation}\label{eq:SR} 
  E(\underline{x}_{H})=\frac{1}{2}(\Vert\underline{x}_{L}-K\underline{x}_{H}\Vert^2_2+\lambda f(\underline{x}_{H}))
\end{equation}
where the last regularization term, which is needed for the problem to be well-posed, is a given norm of the image gradient. The minimization of Eq.~\ref{eq:SR} is found iteratively starting from a first guess on ${x}_{H}$, e.g. a bilinear or polinomial interpolation of ${x}_{L}$. We have designed a FORTRAN90 code (\verb|SuperResolve|) to implement variational super-resolution reconstruction of astronomical images based on L$^1$ and L$^2$ regularization following \citet{unger} and \citet{zomet} respectively. 

\textbf{Tests on simulated EUCLID images.} The Euclid satellite will observe $>$15000 sq. deg. of the sky with a visible (VIS) and near-infrared imager (NIR) to constrain cosmological parameters to an unprecedented accuracy \citep{redbook}. SR reconstruction can be of particular interest to match the resolution (sampling) of VIS (pix-scale=0.1", PSF-FWHM=0.2") and NIR (pix-scale=0.3", PSF-FWHM=0.3") images thus improving reliability of photometric analysis.
Through a dedicated simulation code we have  produced simulated VIS and NIR frames by degrading existing HST-CANDELS images. We applied SR-L2 and SR-L1 reconstruction to a set of 9 NIR single epoch frames to produce a super-resolved mosaic at the same pixel sampling (0.1") of the VIS mosaic.
As shown in Fig.~\ref{fig3} SR-mosaic show an higher level of details, and the PSF-FWHM measured on them is not as worsened with respect to the real value as in the case of pixel upsamling through interpolation.
The SR-L2 technique is particularly promising since fluxes of the sources are preserved and the final S/N reached is the same as expected in a standard coadded mosaic.
\bibliography{O8-3.bib}

\end{document}